\documentclass[12pt]{article}
\usepackage{latexsym}

\def\cH{{\cal H}}
\def\cN{{\cal N}}
\def\cD{{\cal D}}

\def\tr{{\rm tr}}
\def\ket#1{\mid~\!\!\!{#1}~\!\!\rangle}
\def\bra#1{\langle~\!\!{#1}~\!\!\!\mid}

\def\mi{measuring instrument}

\def\QM{quantum mechanics }
\def\qm{quantum mechanics}
\def\QMl{quantum-mechanical }
\def\qml{quantum-mechanical}
\def\CC{calibration condition }

\def\PRC{probability reproducibility condition }
\def\prc{probability reproducibility condition}

\def\${\enskip$}
\def\M{measurement }
\def\m{measurement}
\def\Q{quantum }

\def\IN{^\mathbf i}
\def\f{^\mathbf f}

\begin{document}

\begin{center} {\LARGE \bf Does Unitary Measurement Theory\\ Lead to an Everettian Interpretation?} \vspace{.3cm}

\bf \large  Fedor Herbut\\
\end{center}

\noindent {\footnotesize \it Serbian Academy
of Sciences and Arts, Serbia, Belgrade, Knez
Mihajlova 35}

\vspace{0.3cm} \noindent \rule{13.6cm}{.4pt}

\noindent {\bf Abstract}\\
Quantum-mechanical interpretation-related implications of the theory of unitary pre\M \cite{BLM} on complete \m (objectification or collapse included) are investigated in the present article with a view to give an affirmative answer to the question in the title. It is argued that both Bohr's and von Neumann's ideas lead to those of Everett. Hence, the latter can be, in some sense, considered to be a continuation and elaboration of both former approaches. The importance of the idea of relativeness in Everett's theory is emphasized. To free the relative-state theory from its roots both of classicalness and of subjective observation in the argument of this study, the general or unfolded version of Everett's theory is sketched.\\
\noindent \rule{13.6cm}{.4pt}


\vspace{0.3cm} \normalsize \rm

The truth about physical objects must be strange. It may be unattainable, but if any philosopher believes that he has attained it, the fact that what he offers as the truth is strange ought not to be made a ground of objection to his opinion.

Bertrand Russell
\section {Introduction}

In this study the relative-state interpretation of \qm , initiated by Everett \cite{Everett1}, \cite{Everett2} and further developed by De Witt \cite{De Witt1}, \cite{De Witt2} and by Vaidman \cite{Vaidman1}, \cite{Vaidman2} (and in the present volume) into the many-worlds theory (see also \cite{newBOOK} and \cite{Wallace}), is approached from a somewhat different angle. (I made a previous attempt \cite{FHRC} at a relative-state study formally independent of Everett's theory.)\\
\vspace{0.2cm} \noindent \rule{5cm}{.4pt}

fedorh@sanu.ac.rs

In section 2, a detailed argument is given to the effect that Bohr's insistence on classicalness of the \mi s and von Neumnn's idea of subjective observation both can be viewed to lead to Everett's relative-state interpretation of \qm .

In section 3 the results of the preceding section are discussed in more detail (in three subsections). To free the relative-state theory from its presented roots of classicalness and of subjective observations, which both take the form of ideal \Q pre\m , the latter is replaced by general exact pre\M in section 4. The generalizations represent 'reading of the results' or, equivalently (but sounding more objectively)   'processing the information in the entanglement' in the sense of complete \m .

It turns out that a form of Everett's formalism is required that is wider than the usual one. Therefore the formalism is 'unfolded' in section 5. Section 6 presents 8 final remarks.

The required elements of unitary pre\M theory are sketched in Appendix A leaning on relevant previous
work \cite{FHMeasAlg} and \cite{FHSubsMeas}.

\normalsize
\noindent
The contents of this article reads:

 2\enskip\enskip From Bohr and von Neumann to Everett

3 \enskip\enskip Discussion

3.1 More on Von Neumann, Bohr and Everett

3.2 'Relativeness' and 'absoluteness'

3.3 Critical remarks and possible extensions for remedy

4\enskip\enskip The General Case of 'Reading the Result'

5  \enskip\enskip Unfolding the Formalism of Relative-State Quantum Mechanics

5.1 Relation between relative-state and standard approaches

5.2 Unfolding the relative-state formalism

5.3 Improper mixtures

6   \enskip\enskip Concluding Remarks

 App A  Elements of Unitary Premeasurement Theory

App B Proof of equivalence of the
      partial-scalar product and the
      partial-

      \enskip\enskip \enskip\enskip \enskip\enskip \enskip trace relations

App C Generalization from ideal to
      general exact measurement\\

\section{From Bohr and Von Neumann to Everett}

The usual idea of the so-called {\bf paradox of \M } in \QM is that, if one forgets about classical physics, one lacks a unitary theory of complete \M that would answer the following {\bf basic questions}, as it is required by experience with \Q \m :

{\bf (i)} {\bf Question on decoherence} How does it come about that the {\bf coherence} between the terms in the decomposition of the final pre\M state vector into complete-\M terms (cf (A.1) and (A.4a) in Appendix A) {\bf is lost}?

{\bf (ii)} {\bf Question on selection} How is it possible that the experimenter sees only one of the terms in the mentioned complete-\M expansion of the final pre\M state vector? (We come back to these questions in relation (4) in subsection 3.2 .)\\

Our experience with \Q \M is not very different from that with classical \m . Hence no wonder that the above questions are put in a classical, or at best, quasi-classical way, expecting quasi-classical answers (cf relation (4) below), which the \Q formalism cannot give (cf the no-go theorem of von Neumann \cite{vNeum}, last section VI.3 there).\\

Quasi-classical expectations have often proved no more than a prejudice, i. e., an obstacle that we ourselves put unwittingly in the way of progress. Having in mind this warning, we generalize the questions allowing the \Q formalism to find, possibly, its own answers.

{\bf The generalized questions.}

(i) {\bf Decoherence} Can the \Q formalism furnish decoherence of the complete-\M results at the end of \M in some way?

(ii) {\bf Selection} If the answer to question (i) is affirmative, how does selection of one complete-\M result in the case of individual systems come about?\\

The \QMl formalism is not derived from some more fundamental principles; it comes from experience. Hence it is reasonable to follow Niels {\bf Bohr} and take into account the experimental fact that the {\bf macroscopic, classical measuring instruments obey classical laws}. In particular, this immediately converts pre\M theory into complete \m .

This point has been repeatedly emphasized in the literature, see e. g. the first passage of the last chapter of the book on quantum mechanics by Asher Peres \cite{Peres}. He introduced the catchy term 'dequantization'.

To obtain a rough idea of dequantization, one must realize that once the 'pointer positions' \$\{F_B^k:\forall k\}\$ (cf Appendix A) are classical events, then the {\bf basic law of classical events} is valid for them. It claims that for any sum of disjoint events like \$\sum_kF_B^k=I_B\$ that amounts to the certain event \$I_B\$ {\bf precisely one} of the event terms is bound to occur on the individual system.

We can translate back this classical view into \qm , or to 'requantize' it (to use a term inspired by the mentioned ideas of Peres). Quantum occurrence takes place in \m ; classical happening does not. But since classical ideal \M does not change the state at all, we can assume that the latter {\bf undergoes classical ideal \m . This is best translated into \QM by taking resort to \QMl ideal pre\M }, which at least does not change the state when the initial state has a sharp values of the measured observable (cf (A.6c)).\\

On the other hand, in our search to find answers to the above questions, we can study the case of an arbitrary pre\m , followed by 'reading' the result on part of a sentient being, which we denote as subsystem C. This means taking a {\bf von Neumann} chain of two links. Let \$t_\textbf{i}<t_1\f\$ be the initial moment and the final moment of the first pre\m . The latter is at the same time the initial moment of the second pre\m . Let \$t_2\f\enskip\Big(>t_1\f\Big)\$ be the final moment of the second pre\m .

We have here an example of von Neumann's psycho-physical parallelism \cite{vNeum} (cf p. 421 in section 1. of chapter VI  and p. 440 in the middle of section 3. of chapter VI there) claiming that one can arbitrarily choose which system will be the subject in a von Neumann chain. In our example it is subsystem C.

As it was stated, 'reading' the pointer observable on the measuring instrument on part of the experimenter (subsystem C) can be considered as a classical ideal \m , which does not change any state. It \qml ly becomes ideal \QMl pre\M because the latter is the kind of \Q pre\M that is closest to the former. Hence, we take for the  {\bf second link ideal pre\M }. We express this in more detail.\\

Let $$\ket{\Phi}_{AB}\f=U(t_1\f-t_\textbf{i})_{AB}
\Big(\ket{\phi}_A\IN\ket{\phi ,t_\textbf{i}}_B\Big)=\sum_kF_B^k\ket{\Phi}_{AB}\f
\eqno{(1)}$$ be the final state of the first pre\m ,
i. e., the first link in the chain. Here \$U(t_1\f-t_\textbf{i})_{AB}\$ is the unitary evolution operator of the \M (which includes the object-measuring-instrument interaction), \$\ket{\phi}_A\IN\$ is the initial state of the object, and \$\ket{\phi ,t_\textbf{i}}_B\$ is the initial or "ready-to-measure" state of the instrument. The decomposition into complete-\M terms in the last expression is obtained utilizing the completeness relation \$\sum_kF_B^k=I_B\$, where \$I_B\$ is the identity operator for the instrument.
(If the reader has difficulty with the notations, perhaps he should read Appendix A first.)

The second pre\M 'reads' the first-pre\M result. As it has been explained above, it is an ideal pre\M . It is the pointer observable \$P_B=\sum_kp_kF_B^k\$ of the first pre\M that is the measured observable in it.

As to the second link, we utilize the fact that,
according to relation (A.6a), ideal pre\M leaves unchanged each of the selective  components \$F_B^k\ket{\Phi}_{AB}\f\$ in expansion (1), but it multiplies tensorically each of them with a different new 'pointr position' state \$\ket{\phi ,t_2\f}_C^k\$. These states are orthogonal to each other. Altogether,
$$\ket{\Psi}_{ABC}\f=U(t_2\f-t_1\f)_{ABC}\Big(
\ket{\Phi}_{AB}\f\ket{\phi,t_1\f}_C\IN\Big)=$$
$$\sum_kF_B^k\ket{\Phi}_{AB}\f
\otimes \ket{\phi , t_2\f}_C^k.\eqno{(2)}$$\\

Next, we evaluate a relevant decomposition of the final subsystem state \$\rho_{AB}\f\equiv\tr_C\Big(\ket{\Psi}_{ABC}\f
\bra{\Psi}_{ABC}\f\Big)\$, where "$\tr_C$" denotes the partial trace over subsystem C. We turn to the general \Q formalism and do this with the help of an {\bf auxiliary claim}, which we supply with physical meaning in the standard (Copenhagen) way:\\

Let\$\ket{\Psi}_{12}\$ be any bipartite state vector, and \$\{\ket{n}_2:\forall n\}\$ a complete orthonormal basis for the second subsystem.
Let, further, $$\ket{\Psi}_{12}=\sum_n
\overline{\ket{n}}_1\ket{n}_2\eqno{(aux-1 )}$$ be the expansion of the former in the latter. (If unfamiliar with expansion in a subsystem basis, cf section 2 in \cite{FHArxSchmidt}.) The 'expansion coefficients'
\$\{\overline{\ket{n}}_1:\forall n\}\$ are vectors in the state space of the opposite subsystem. They are not necessarily unit vectors (to indicate this, they are written overlined).

Then, one immediately obtains
$$\rho_1\equiv\tr_2\Big(\ket{\Psi}_{12}
\bra{\Psi}_{12}\Big)=$$ $$\sum_n||\overline{\ket{n}}_1||^2
\times\Big(\overline{\ket{n}}_1\Big/
||\overline{\ket{n}}_1||\Big)\Big(\overline{\bra{n}}_1
\Big/||\overline{\ket{n}}_1||\Big).\eqno{(aux-2)}$$ It is an {\bf improper mixture} \cite{D'Esp} (subsection 7.2 there). The {\bf statistical weights}
\$\{||\overline{\ket{n}}_1||^2:\forall n\}\$ in it are the {\bf probabilities} of the 'occurrences' of the corresponding events \$\{\ket{n}_2\bra{n}_2:\forall n\}\$ in the composite state \$\ket{\Psi}_{12}\$. The pure states $$\forall n:\quad\overline{\ket{n}}_1\Big/
||\overline{\ket{n}}_1||\eqno{(aux-3)}$$
are the {\bf 'conditional states'} for which the corresponding {\bf conditions} are the 'occurrences' of the corresponding events \$\{\ket{n}_2\bra{n}_2:\forall n\}\$ in
\$\ket{\Psi}_{12}\$.\\

In order to return to our case given by relation (2), we note that our subsystems are A+B (subsystem 1) and C (subsystem 2). We can complete, if necessary,  the sub-basis \$\{\ket{\phi ,t_2\f}_C^k:\forall k\}\$ into a basis for subsystem C in an arbitrary way. Relation (2) is the expansion of our bipartite state in this basis. (The coefficients along the basis vectors that are joined to the former sub-basis are all zero.)

One can see now that application of the general auxiliary decomposition (aux-1-aux-3)to the composite-system state given by (2) gives:

$$\rho_{AB}\equiv\tr_C\Big(\ket{\Psi}_{ABC}\f
\bra{\Psi}_{ABC}\f\Big)=
\sum_k||F_B^k\ket{\Phi ,t_1\f}_{AB}||^2\times$$  $$  \Big(F_B^k\ket{\Phi ,t_1\f}_{AB}\Big/
||F_B^k\ket{\Phi ,t_1\f}_{AB}||\Big)
\Big(\bra{\Phi ,t_1\f}_{AB}F_B^k\Big/
||F_B^k\ket{\Phi ,t_1\f}_{AB}||\Big). \eqno{(3)}$$\\

The improper mixture (3) can provide us with {\bf answers} to the two generalized questions aimed at complete \m s above if we make two {\bf stipulations}.

(a) The experimenter or observer is quantum-mechanically just a measuring instrument with the relevant contents of his consciousness as the 'pointer positions' \$F_C^k\$. Hence, relation (2) is the relevant all-encompassing state.

(b) We do not stick to the idea of 'absoluteness' in \QMl description, and allow for 'relativeness'. This is due to the above conditional states, which have physical meaning relative to the above conditions.

It is noteworthy that the first two 'questions' at the beginning of the section were stated with the underlying idea of 'absoluteness' suggested by the quasi-classical approach (cf relation (4) below in subsection 3.2). One should also note that we are dealing with 'relativeness' in non-relativistic \qm .\\

The \QMl formalism, as it has been developed so far, suggests the following answer to the generalized questions.

(i) We have seen in (3) that {\bf relative to subsystem A+B} of the object (A) plus first measuring instrument (B), the coherence has disappeared; {\bf decoherence} has set in.

(ii) If we define a \M result in the first pre\M {\bf relative to} the second measuring instrument (subsystem C, the experimenter) plus one of its 'pointer positions' \$F_C^{\bar k}=\ket{\phi , t_2\f}_C^{\bar k}\bra{\phi , t_2\f}_C^{\bar k}\$ (the contents of the consciousness of the experimenter observing the result of the first pre\M ), then we do have complete \m .\\

Our 'relative answers' are precisely what Everett has brought as novelty into the interpretation of \QM \cite{Everett1}, \cite{Everett2}. His approach differs from the standard, i. e., Copenhagen one in several aspects. The conditional states are, according to Everett, {\bf relative states}; and they are not 'on condition that the event \$\ket{n}_2\bra{n}_2\$ occurs' (see the auxiliary claim above),  but, instead, they are {\bf 'in relation to the state} \$\ket{n}_2\$'. These may be viewed as semantic differences.

Where Everett differs most and substantially from the Copenhagen view is in the assertion that {\bf the  pre\M final state vector} \$\ket{\Psi}_{ABC}\f\$ {\bf 'remains fully preserved' in complete \m }. The 'relativeness' in the approach makes this possible.\\

\section{Discussion}

Everett's approach to complete \m , as we have obtained it in the preceding section, requires critical examination. We have not {\it derived} 'relativeness' and Everett's view from unitary \Q pre\M theory; we have only recognized that it is a possible answer to the generalized questions concerning complete \m .

What is left open? We certainly cannot get answers as expected in the formulation of the  traditional, quasi-classical questions. Logically it is possible that some new insight might get answers to the generalized questions differing from the Everettian ones. I do not think that this is likely to happen, not within unitary dynamics. (See also concluding remark (ii) in the last section.)\\

It seems desirable to relate our conclusions in the preceding section to known facts and opinions from the literature.\\

\subsection{More on Von Neumann, Bohr and Everett}

Von Neumann (or the 'orthodox interpretation' of \QM as contrasted with the Copenhagen one) applies the {\bf projection postulate} besides unitary theory, projecting out all but one complete-\M component
\$F_B^{\bar k}\ket{\Phi}_{AB}\f\Big/||F_B^{\bar k}\ket{\Phi}_{AB}\f||\$
of the final pre\M state vector \$\ket{\Phi}_{AB}\f\$ (cf (1)). He called it "process one", whereas the unitary evolution was "process two". But, it was Niels Bohr who played the most outstanding role in viewing \QM in the framework of 'absoluteness'.\\

I'll argue that {\bf Bohr} (the father of the Copenhagen interpretation) and {\bf Everett need not be considered as violently conflicting views}. For a start, I turn to Shimony's incisive analysis of Bohr \cite{Shimony} (pp.769 and 770 there).

"I suspect that Bohr was aware of the difficulties inherent in a macroscopical ontology, and in his most careful writing he states subtle qualifications concerning states of macroscopic objects. For example" \cite{Bohr}

\begin{quote}
"The main point here is the distinction between the {\it objects} under investigation and the {\it measuring instruments} which serve to define, in classical terms, the conditions under which the phenomena appear. Incidentally, we remark that, for the illustration of the preceding considerations, it is not relevant that experiments involving an accurate control of the momentum or energy  transfer from atomic particles to heavy bodies like diaphragms and shutters would be very difficult to perform, if practicable at all. It is only decisive that, in contrast to the proper measuring instruments, these bodies together with the particles would constitute the system to which the \QMl formalism has to be applied."
\end{quote}

Shimony further comments: "Bohr is saying that from one point of view the apparatus is described classically and from another, mutually exclusive point of view, it is described \qml ly. In other words, he is applying the principle of complementarity, which was originally formulated for microscopical phenomena, to a macroscopic piece of apparatus. ..."\\

It seems to me that in Bohr's macroscopic complementarity principle (as we perhaps may call it following Shimony's discussion) there is concealed an important implication. If the \QMl formalism can be applied to
"bodies" like diaphragms and shutters, then it can, most likely, be applied to all classical systems. But then, clearly, Bohr admits that macroscopic systems, usually described classically, can be described also \qml ly. This opens up the alley of treating the macroscopic objects on the same footing as the microscopic ones (cf \cite{Cooper}, \cite{FHScully1}, \cite{FHScully2}).

Thus, seeds of Everett's approach, who assumed that there is no limit to the applicability of \qm , are present already in the "most careful writing" of Bohr (as Shimony puts it).\\

As it was stated, Everett's most provocative claim for the adherents to the Copenhagen interpretation is the claim that the final state \$\ket{\Phi}_{AB}\f\$ does not split up into the final complete-\M components abolishing coherence (more on this in subsection 3.2 and relation (4)). Decoherence in the subsystem A+B, with preservation of coherence in the larger, comprehending system A+B+C in the state \$\ket{\Psi ,t_2\f}_{ABC}\$, is, surprisingly or not, in some way, furnished by unitary theory (in the preceding section), which appears to be in no actual conflict with the Copenhagen view of \qm .

One should note that elevation of the coherence by interaction to the larger, comprehending system (ABC in (2)) accompanied by decoherence in the subsystem (A+B) (cf (3)) is an important fundamental theoretical fact inherent in von Neumann's writings \cite{vNeum}, emphasized in \cite{ZehJoos85}, and discussed in my previous work \cite{FHScully1}.\\

It seems to me that the crucial point in comparing Bohr and Everett is seeking answer to the question if 'survival' of the other branches, complete-\M components, in which the experimenter does not find himself, has any physical meaning. I suspect that it does.

We have reached Everett's view by stipulating that a sentient observer is not different from an inanimate measuring instrument as far as the \Q formalism is concerned. Analogously, the formalism makes no distinction between small, microscopic systems and large macroscopic, classical ones. Let us take a microscopic example.

Suppose we measure the spin projection of  a spin-one-half particle in the Stern-Gerlach (SG) measuring instrument (the spin tensor-factor space of the particle is the object subsystem A). Let us forget about the screen for a moment. Spin projection and moving into the upper or lower half-space in the SG magnetic field
satisfy the \CC (cf Appendix A): if spin is 'up', the particle moves upwards,if spin is 'down', the particle moves downwards. Thus, the orbital (spatial) tensor factor in the state vector of the particle in the magnetic field fully qualifies as a measuring instrument in the formalism (subsystem B), and the final pre\M state \$\ket{\Phi ,t_1\f}_{AB}\$ is the
 state vector of the particle when leaving the SG magnetic field (before it hits the screen).

Does one of the upper-half-space and lower-half-space components of this final state disappear in the field? The \Q formalism implies and we have reason to believe that the disappearance of one of the components does not take place earlier than when the particle hits the screen. This is confirmed by the relevant analogous neutron interference experiments, e. g., \cite{Rauch}.

When the particle in the magnetic field of the Stern-Gerlach measuring instrument hits the screen, only one of the two final complete-\M components is observed. The other component seems to vanish completely. This fact, and the same phenomenon in numerous variations of  concrete \m s, have puzzled and haunted many great thinkers in physics. This is the core of the paradox of \Q \M theory. No wonder that von Neumann and essentially also Bohr discarded by postulate all but the observed final complete-\M component (one term in the mixture (3)). It came as a reasonable, but essentially phenomenological, postulate. Their attitude was within the framework of 'absoluteness'. There appeared to be no alternative.

Everett payed a high price in personal suffering \cite{EverettByrnearticle} \cite{EverettByrneBOOK} for not taking the reasonable path. Instead, he followed consistently the \Q formalism (perhaps believing blindly that it contains the underlying \Q laws of nature). Thus he found the obvious alternative within the frame of 'relativeness'.

Returning to the Stern-Gerlach \m , before the particle reaches the screen, we have the final state \$\ket{\Phi ,t_1\f}_{AB}\$ of the first pre\m , consisting of the mentioned two terms. The spot on the screen is the result of the second \m , which is a complete one.

In the Everettian view, the first pre\M took place in the laboratory, i. e. the experimenter was faced with the mentioned coherent mixture of complete results analogously as 'Wigner' was in relation to his 'friend' in the famous paradox of 'Wigner's friend' \cite{Wigner}. This {\bf changes drastically} in the second pre\M , when the experimenter, all his macroscopic devices and the whole laboratory, join one of the former complete-\M components.

In the Everettian view, the drastic change takes place when the consciousness of the experimenter becomes one of the states \$\ket{\phi ,t_2\f}_C^{\bar k}\$ for a fixed value \$\bar k\$ in the expansion (2). Then the consciousness, together with the entire body of the experimenter, even the whole laboratory and its surroundings, acquire a double identity: they still are a tensor factor in the tensor product of different subsystems (preserving their basic physical identity), and they belong now to an Everettian branch (denoted by \$\bar k\$ for instance). The other final complete-\M components do not disappear; they are just not present in the branch  corresponding to \$\bar k\$.

In the Copenhagen approach one emphasizes the difference between macroscopic and microscopic. It is only the former where classical physics, which lacks the very notion of coherence, can be applied.

In the Everettian approach one would say that, on account of the complexity of macroscopic objects, it is hard, but, in principle, not impossible to see interference of two or more complete-\M components in the laboratory.

The present author is inclined to consider the possibility that the  {\bf Everettian approach might be a successor of and improvement on the Copenhagen approach}. Future results will tell.\\

\subsection{'Relativeness' and 'absoluteness'}

To avoid the term 'relativity', which is somehow reserved for Einstein's special and general relativities, the term 'relativeness' is used in this article for relative concepts within non-relativistic \qm . 'Absoluteness' is used as the negation of 'relativeness'.

Classical physics is permeated with the hidden idea of 'absoluteness'. It is hidden because in classical physics there is no alternative; it is self-evident. It seems to me that a simple definition of 'absolute' goes as follows.

Let us have a physical system A, and let us make a physical statement about it. Further, let B be another physical system having no overlap with A. The larger system A+B contains A as a subsystem. If we can refute the mentioned statement in A+B, then it is not absolute; it is valid only relative to subsystem A. If, on the other hand, it is true that for every subsystem B the statement is equivalently valid in A+B, then it is valid in an absolute sense.

In the particular example of 'state' of a system A, classically, a state is either pure or mixed, and this cannot be proved false in any larger system containing A; it is an absolute statement. I believe, nobody says so, because all physical statements are such in classical physics. What is more, one might formulate the classical criterion that a physical statement is true only if it is equivalently valid in every larger system containing the initial one.

In \QM we have proper and improper mixtures \cite{D'Esp} (subsection 7.2 there). This is an important case where the distinction between 'absoluteness' and 'relativeness' appears.

Let us start with {\bf absoluteness}. The formulation of the quasi-classical questions concerning complete \M (cf beginning of section 2) was in hope of a quasi-classical answer that would, in view of the coherence of the terms in expansion (1), read
$$\ket{\Phi}_{AB}\f\quad\rightarrow\quad\rho_{AB} \f
\equiv$$  $$\sum_k\bra{\phi}_A\IN E_A^k\ket{\phi}_A\IN\times\Big(
F_B^k\ket{\Phi}_{AB}\f\Big/||F_B^k\ket{\Phi}_{AB}\f ||\Big)\Big(\bra{\Phi}_{AB}\f F_B^k \Big/||F_B^k\ket{\Phi}_{AB}\f ||\Big)\eqno{(4)}$$ to obtain the relevant mixture of complete-\M final states. It is called a {\bf proper mixture}.

We are concerned with the 'absoluteness' of the proper mixture (4). If we consider a larger system by joining any other system C in some state \$\rho_C\$, then the transition from pre\M to complete \M is $$\Big(\ket{\Phi}_{AB}\f \bra{\Phi}_{AB}\f\Big)\otimes\rho_C\quad\rightarrow
\quad\rho_{AB}\f\otimes\rho_C,$$ and subsequent evaluation of the final complete-\M AB-subsystem state brings us obviously back to \$\rho_{AB}\f\$ in (4). The point is, of course, the lack of entanglement between AB and C.

Contrariwise, if we have another \M (for instance, an ideal one as we had so far), then suitable entanglement is created between subsystems AB and C,  and a decomposition identical to (3), but this time with a meaning relative to subsystem AB only (and the terms having meaning relative to the distinct pointer positions of \$F_C\$) comes about. We thus obtain a so-called {\bf improper mixture} \cite{D'Esp} (subsection 7.2 there) with the described relativeness.\\

Let me close this section with a short story, which might appear naive at first glance; but actually it is profound. A boy is standing on the riverbank. Somebody is shouting to him from the opposite bank: "Boy, how can I go to the other bank?" "You fool" - answers the boy, "You are already on the other bank."

When the boy grows up, he will realize that "this side" and "the other side" are relative concepts in relation to the speaker. It could be that Everett's ideas are prompting us to grow up in comprehending \qm .\\

\subsection{Critical remarks and possible extensions\\ for remedy}

Let us start this subsection with some {\bf critical remarks}.

The tacit assumption that the system A+B (object plus measuring instrument) can be dynamically isolated from its surroundings during pre\M (from \$t_\textbf{i}\$ till \$t_1\f\$), may raise suspicion; in reality it might be illusionary. So is the underlying tacit assumption of statistical isolation, i. e., that the subsystems considered have not interacted with the surroundings in the past, so that there are no statistical correlations between the former and the latter or that in preparation such correlations, if they existed, could have been destroyed.

The preparator may play an important role in achieving pure-state objects and pure initial states  of the measuring instrument. These states can be conditional states, in the standard language of Copenhagen, the conditions being some events on the preparator, or relative states relative to some preparator and events in the terminology of Everett.\\

One wonders if there is hope of improvement. In the entire unitary theory of \Q pre\M (cf the review \cite{FHMeasAlg}) the very definition of subsystem B is to some extent incomplete. Subsystem B can contain parts that have no connection with the process of pre\M .

Taking into account the above critical thoughts, we may try to improve the theory. But first, a finer analysis is required.

We start with the fact that in macroscopic measuring devices usually only a part (a subsystem) performs the pre\M . Hence, we may view the instrument B as consisting of two disjoint parts, \$B=B_1+B_2\$, where only \$B_1\$ takes part in the measuring process. (\$B_2\$ may interact with the object A and with \$B_1\$, but it is irrelevant for the pre\M .) We may define \$B_1\$ to be {\bf minimal}, i. e., so that the pointer observable, as an operator, is actually acting in the state space of subsystem \$B_1\$, and not in the state space of any subsystem of \$B_1\$. In other words, \$B_2\$ is the largest 'passive' part of B (meaning 'relative to pre\m ').

In our description of pre\M , subsystem \$B_2\$ need not be confined to part of the actual measuring device; it may encompass a part of the surroundings; actually, as much of it as it takes to invalidate the above critical remarks on dynamical and correlational isolation.

In view of the fact that all bodies in the universe seem to interact all the time with each other, the safest thing may be including all the rest of the world in \$B_2\$. We will see in the next section what this means in more detail.\\

\section{The General Case of 'Reading the Result'}

It seems reasonable to investigate if, within the unitary formalism of \qm , ideal pre\M is replaceable by general exact pre\m . Namely, in \Q pre\M theory the former is often a springboard for the latter (cf e. g. \cite{FHSubsMeas}); one may suspect that ideal pre\M  is not endowed with an important universal role.

If the second pre\M in the two-link von Neumann chain that we have discussed in section 2 is not an ideal pre\m , then we lack a sub-basis the elements of which are distinguished by distinct pointer positions, referring to different results, which led to an Everettian interpretation in a simple way.\\

We have made the stipulation that the living observer is, as far as \M theory in \QM is concerned, just a measuring instrument (subsystem C). As a consequence, we have seen that his \m -theoretic 'processing' of the information contained in the \Q correlations created by the second pre\M led to decoherence in the subsystem state \$\rho_{AB}\$ and to complete \M  in a relative sense (cf relation (3)). (This "processing" was previously called subjectively "reading the result".)

The question arises if every second exact pre\M  can, analogously, process the relevant \m -theoretic information from the entanglement that came about in the second pre\M .\\

We again turn to the \Q formalism for an affirmative answer. One has the following relevant {\bf general auxiliary claim}.\\

Let \$\rho_{12}\$ be a general composite state (density operator), and let  \$\sum_nP_2^n=I_2\$ be an arbitrary (finite or infinite) {\it decomposition of the identity}, i. e., an orthogonal sum of projectors adding up to the identity operator, for subsystem 2. Then, the first-subsystem state (reduced density operator) \$\rho_1\equiv\tr_2\rho_{12}\$
can be written in a decomposed form as follows:
$$\rho_1=\sum_nw_n\rho_1^n,\eqno{(AUX-a)}$$ where $$\forall n:\quad w_n\equiv\tr(\rho_{12}P_2^n),\eqno{(AUX-b)}$$ and  $$\forall n,\enskip w_n>0:\quad\rho_1^n\equiv
\tr_2(\rho_{12}P_2^n)\Big/\tr(\rho_{12}P_2^n).
\eqno{(AUX-c)}$$\\

{\it Proof} One can write $$\rho_1\equiv\tr_2\rho_{12}=\sum_n\tr_2
(\rho_{12}P_2^n),$$
from which the claimed decomposition AUX-a with the relations AUX-b and AUX-c immediately follows. {\it This ends the proof.}\\

{\bf Physical meaning} of the entities in the standard (non-Everettian) sense: The decomposition (AUX-a)-(AUX-c) itself appears to be a {\bf mixture}. The statistical weights \$\{w_n:\forall n\}\$ are the {\bf probabilities} of occurrence of the corresponding events \$\{P_2^n:\forall n\}\$ in the state \$\rho_{12}\$.

Each subsystem state \$\{\rho_1^n:\forall n\}\$ is the {\bf conditional state} corresponding to the {\bf condition} that the event \$P_2^n\$ {\bf occurs} in the state \$\rho_{12}\$. Namely, utilizing idempotency and under-the-partial trace commutativity (cf (A.7)), one can rewrite (AUX-c) as

$$\forall n,\enskip w_n>0:\quad\rho_1^n=
\tr_2(P_2^n\rho_{12}P_2^n)\Big/\tr(P_2^n\rho_{12}P_2^n).
\eqno{(AUX-d)}$$

An event \$P_2^n\$ 'occurs' in \QM in complete \M of the observable \$P_2^n=1\times P_2^n+0\times (P_2^ n)^c\$ ((\$P_2^ n)^c\$  is the complementary event) corresponding to the result \$1\$ (and one shortly says '$P_2^n$ is measured'). At first glance, as one can see in relation (AUX-d), the pre\M in question is an ideal one (cf (A.6b)). But, if one takes into account the result obtained in \cite{FHSubsMeas}, which says that as far as the effect on the opposite subsystem is concerned, all pre\m s have the same effect as ideal pre\M , then it is clear that the kind of pre\M is immaterial. (In the mentioned reference the claim was proved only for pure composite-system states. But it is easily generalized into the analogous claim for general states - cf beginning of Appendix C.)\\

We can utilize the auxiliary general claim to answer the question of processing the relevant information  as follows. Our composite state is \$\ket{\Psi}_{ABC}\f\$ (cf (2)), and it is the completeness relation \$\sum_kF_C^k=I_C\$, accompanying the pointer observable \$P_C=\sum_kp'_kF_C^k\$, that is the relevant decomposition of the identity. Relation (AUX-a) then becomes $$\rho_{AB}\f
\equiv\tr_C\Big(\ket{\Psi}_{ABC}\f\bra{\Psi}_{ABC}\f
\Big)
=\sum_k\bra{\phi}_A\IN E_A^k\ket{\phi}_A\IN\times
\rho_{AB}^{\textbf{f},k}\eqno{(5a)}$$ (cf (2) and (A.4a)), where, due to (AUX-c), \$\forall k,\enskip
\bra{\phi}_A\IN E_A^k\ket{\phi}_A\IN >0$:
$$\rho_{AB}^{\textbf{f},k}\equiv\tr_C\Big(
\ket{\Psi}_{ABC}\f
\bra{\Psi}_{ABC}\f F_C^k\Big)\Big/\Big[\tr\Big(
\ket{\Psi}_{ABC}\f
\bra{\Psi}_{ABC}\f F_C^k\Big)\Big].\eqno{(5b)}$$

Since we have a pure state \$\ket{\Psi}_{ABC}\$ of the larger comprehending system, (5a) cannot be an ordinary, i. e., absolute  mixture, and therefore (5a) is an {\bf improper mixture}.

Applying the idea of 'relativeness', we can say that the mixture in (5a), which displays decoherence, is valid relative to subsystem AB, and that, for
\$\forall k,\enskip\bra{\phi}_A\IN E_A^k
\ket{\phi}_A\IN >0\$, the subsystem state \$\rho_{AB}^{f,k}\$ is the state relative to the event ('pointer position') \$F_C^k\$ in the (unchanged) composite-system state \$\ket{\Psi}_{ABC}\$.

We have thus obtained a {\bf relative-state physical meaning} of the processing of the entanglement between subsystems \$A+B\$ and \$C\$
in the final state \$\ket{\Psi}_{ABC}\f\$ in the spirit of Everettian ideas. Subsystem C thus "reads" the \M results in the first pre\m .\\

One should note that the described processing of relevant information brings us back to von Neumnann's psycho-physical parallelism (cf section 2) with an arbitrary (subjective) cut between object and subject.
In (5a) the object of physical description is subsystem \$A+B\$. The first measuring instrument with the 'pointer positions' \$\{F_B^k:\forall k\}\$ belongs to the object of physical description.\\

We return now to the last point in the preceding subsection, where \$A+B_1+B_2\$ constituted the entire world on account of a suitable broadening of \$B_2\$. In the final pre\M state vector \$\ket{\Phi}_{AB}\f\$ (cf(1)) we substitute \$B_1+B_2\$ for \$B\$, and we evaluate the conditional (or relative) states
$$\forall k:\quad \rho_{AB_2}^{\textbf{f},k}
\equiv\tr_{B_1}\Big(\ket{\Phi}_{AB_1B_2}\f
\bra{\Phi}_{AB_1B_2}\f F_{B_1}^k\Big)\Big/
\tr\Big(\ket{\Phi}_{AB_1B_2}\f
\bra{\Phi}_{AB_1B_2}\f F_{B_1}^k\Big).\eqno{(6)}$$

To discuss (6), let us take two extreme cases.

(i) If there is no entanglement between \$A+B_2\$ and \$B_1\$, then the tripartite state vector tensor-factorizes into the state vector of subsystem \$(A+B_2)\$ and that of \$B_1\$. As a consequence,
for all \$k\$ values (for all branches) subsystem \$A+B_2\$ is in one and the same state as one can easily see in (6).

(ii) Contrariwise, if the entanglement between \$B_1\$ and \$A+B_2\$ is strongly dependent on the pointer positions \$F_{B_1}^k\$ (which can be the  verbatim pointer positions or the consciousness of the observer), then we have different state vectors for the entire world. This brings us to De Witt's controversial many worlds \cite{De Witt1}, \cite{De Witt2}.  They are in our description the conditional states (relative states) of subsystem \$A+B_2\$ corresponding to the distinct pointer positions \$F_{B_1}^k\$.\\

\section{Unfolding the Relative-State Formalism}

One may wonder if the answer in terms of relative states {\bf in relation to events} for another system, obtained in the preceding section, is still within the conceptual confines of the Everettian approach. My answer is affirmative. This section is devoted to an explanation of this claim.\\

The general 'processing of the relevant information' or 'reading the result' expounded in the preceding section gives motivation for discussing the Everettian formalism in more detail.

Everett himself in his {\bf relative-state \QM }(RSQM) \cite{Everett1}, \cite{Everett2} used mostly states in relation to {\bf states} of another subsystem. I believe this is the {\bf folded version} of his approach. It can be unfolded in a straightforward way as it will be shown in the following subsections. 'Unfolding' is a generalization in which the 'folded' special case implies, i. e., uniquely determines, the generalization, which is then the 'unfolded' version.\\

\subsection{Relation between relative-state and standard approaches}

The 'folded' version of RSQM has three equivalent forms. In the {\it first}, the relative state \$\ket{\phi}_A\$ of the object that corresponds to the subject state \$\ket{\phi}_B\$ in a pure composite state \$\ket{\Psi}_{AB}\$ is evaluated by the procedure used originally by Everett \cite{Everett1}, \cite{Everett2}. It is the normalized 'expansion coefficient' multiplying in a tensorial way  \$\ket{\phi}_B\$ in the expansion of \$\ket{\Psi}_{AB}\$ in any  basis of subsystem B containing \$\ket{\phi}_B\$ (cf aux-1-aux-3 in section 2).

In the {\it second form} no subsystem basis is required. One just multiplies \$\ket{\Psi}_{AB}\$ from the left by the bra of \$\ket{\phi}_B\$ in a partial scalar product, and then one performs normalization. The equivalence of the first two forms is trivial (cf, if desired, section 2 in \cite{FHArxSchmidt}).

Finally, in the {\it third form} the procedure is further varied, and the partial scalar product is replaced by a suitable partial trace according to the following {\it equivalence}: $$\ket{\phi}_A=\bra{\phi}_B\ket{\Psi}_{AB}
\Big/[\bra{\Psi}_{AB}(\ket{\phi}_B \bra{\phi}_B)\ket{\Psi}_{AB}]^{1/2}
\quad\Leftrightarrow$$
$$\ket{\phi}_A \bra{\phi}_A=\tr_B[(\ket{\Psi}_{AB} \bra{\Psi}_{AB})(\ket{\phi}_B \bra{\phi}_B)]\Big/$$ $$\{\tr[(\ket{\Psi}_{AB} \bra{\Psi}_{AB})(\ket{\phi}_B \bra{\phi}_B)]\}.\eqno{(7)}$$
(Naturally, the implication "$\Leftarrow$" is 'up to an arbitrary phase factor'.) For the  reader's convenience, equivalence (7) is proved in Appendix B.\\

Now we focus attention on the relation between RSQM and the standard approach.
By the latter is meant the simplified, text-book version of the Copenhagen interpretation of \qm .

Following relation (aux-d) in section 2, one can rewrite the partial-trace relation in (7) as follows:
$$\ket{\phi}_A\bra{\phi}_A=$$
$$\tr_B\Big[\Big(\ket{\phi}_B
\bra{\phi}_B\Big)\Big(\ket{\Psi}_{AB}
\bra{\Psi}_{AB}\Big)\Big(\ket{\phi}_B
\bra{\phi}_B\Big)\Big]\Big/$$  $$
\Big\{\tr\Big[\Big(\ket{\phi}_B
\bra{\phi}_B\Big)\Big(\ket{\Psi}_{AB}
\bra{\Psi}_{AB}\Big)\Big(\ket{\phi}_B
\bra{\phi}_B\Big)\Big]\Big\}.\eqno{(8)}$$
In the standard approach LHS(8) is the {\bf conditional state} of subsystem A that is given rise to by the {\bf occurrence} of the opposite-subsystem event \$\ket{\phi}_B
\bra{\phi}_B\$ in the bipartite state \$\ket{\Psi}_{AB}\bra{\Psi}_{AB}\$ in ideal pre\M (cf relation (19b) in \cite{FHSubsMeas}).

This is the {\it connection} between RSQM and the standard approach. They describe the same physical process, the complete ideal \M of the event \$\ket{\phi}_B\bra{\phi}_B\$, in different terms.

As it was stated before (cf end of section 2), but it is worth repeating, it is the pre\M of the same event (as an observable) in the same composite-system state where the two approaches disagree. The standard approach, guided by the classical idea of 'occurrence', discards all other complete-\M final components; whereas RSQM leaves the entire final state of pre\M {\bf unchanged}.

One should note that in RSQM there is no 'occurrence' of events (no collapse). An event plays the role of a {\bf subject entity} in relation to which the relative state \$\ket{\phi}_A\$ is, in a subjective way, considered.

Incidentally, the first form of RSQM, the one most favored by Everett himself, perhaps inspired by von Neumann's maximal over\M of each observable, also the observable \$\ket{\phi}_B\bra{\phi}_B\$ is maximally overmeasured in pre\m . This may lead to infinitely many complete-\M final components, or, equivalently in the language of mathematics, to infinitely many expansions of the composite state in a subsystem basis.\\

\subsection{Unfolding the relative-state formalism}

Having established that it is relation (8) where the two approaches coincide (though using different terms), unique generalization in the \Q formalism is well known. If \$\rho_{AB}\$ is any density operator for the composite system with the physical meaning of a proper mixture (cf subsection 3.2), and \$P_B\$ denotes an arbitrary event (projector) for subsystem B, then the well-known unique generalization of (8) is
$$\rho_A=
\tr_B\Big(P_B\rho_{AB}P_B\Big)\Big/
\tr\Big(\rho_{AB}P_B\Big)\eqno{(9a)}$$ \cite{Lud}, \cite{Messiah}, \cite{Laloe}. For the sake of completeness, this is proved in Appendix C.

In the standard approach, (9a) defines the conditional state of subsystem A due to the occurrence of the event \$P_B\$ on the opposite subsystem in the composite-system state \$\rho_{AB}\$ in ideal pre\M of \$P_B\$. In RSQM (9a) determines the {\bf relative state} of the object subsystem A in relation to the {\bf subject event} \$P_B\$ of the opposite subsystem in the composite-system state \$\rho_{AB}\$.

Going back to the equivalence of relations (AUX-d) and (AUX-c) in section 4, we see that (9a) can be written in the {\bf equivalent} form:
$$\rho_A=
\tr_B\Big(\rho_{AB}P_B\Big)\Big/
\tr\Big(\rho_{AB}P_B\Big).\eqno{(9b)}$$
This is where the unfolded RSQM gives the same relation as the one we have obtained in the preceding section in our analysis of pre\M in (6). The composite-system state was pure there; it was the final state \$\ket{\Phi}_{AB}\f\$ of pre\m , but the subject event was a general event, a 'pointer position' \$F_B^k\$.\\

\subsection{Improper mixtures}

Perhaps it is not superfluous to show that relations (9a) and (9b) are valid also for improper mixtures \$\rho_{AB}\$.

We assume that we have  a general (mixed or pure) state \$\rho_{AB}\enskip\Big(\equiv\tr_C\rho_{ABC}\Big)\$  of a subsystem AB of a tripartite composite system ABC. We assume that the state \$\rho_{ABC}\$ is pure or a proper mixture.

We want to evaluate the relative state \$\rho_A\$ of the first subsystem, which we choose to be our \textbf{object subsystem}. For the \textbf{subject subsystem} we take subsystem B. Finally, we assume that an event (projector) \$P_B\$ on the subject subsystem is given for the \textbf{subject event}. The subject subsystem and its subject event together constitute the \textbf{subject entity}.

We can take subsystem B+C as the subject subsystem with \$P_B\$ as its subject event in an intermediate step. Then
the evaluation of the relative state, according to (9b), goes as follows: $$\rho_A=
\tr_{BC}\Big(\rho_{ABC}P_B\Big)\Big/ \tr\Big(\rho_{ABC}P_B\Big)$$ or, equivalently (as easily seen) $$\rho_A=
\tr_B\Big(\rho_{AB}P_B\Big)\Big/
\tr\Big(\rho_{AB}P_B\Big), \eqno{(10a)}$$
where $$\rho_{AB}\equiv\tr_C(\rho_{ABC})\eqno{(10b)}$$ is the subsystem state (reduced density operator) of the A+B subsystem.

\$\rho_{AB}\$ is a subsystem state, possibly an improper mixture. Relation (10a), defining the relative state \$\rho_A\$, is not different from (9b) that is valid for proper mixtures or pure states. This is consistent with the well-known fact that proper and improper mixtures are equally described in \qm .\\

\section{Concluding Remarks}

{\bf (i)} Though this investigation has not actually derived the Everettian relative-state interpretation from unitary pre\M theory, it has {\bf lent support} to it.\\

{\bf (ii)} One may also argue as follows. If one formulates a {\it principle of impossibility} (analogous to that in thermodynamics)
that one cannot find any other answer to the generalized questions in section 2 than the one presented in section 2, and if the principle would be widely accepted, then the Everettian relative-state interpretation of \QM could be based on it (as the basic laws of thermodynamics are based on the principles of impossibility to create energy out of nothing and to extract work out of heat at a fixed temperature).\\

{\bf (iii)} When I was analyzing a fascinating thought experiment \cite{FHScully1}, \cite{Scully1} and  an even more inspiring real experiment \cite{FHScully2}, \cite{Scully2}, both realizing transition through one of two slits and interference through them in the same experiment (and other wonders), I found that the {\bf relative-state approach} gave an in-depth physical view. But as long as one is confined to analyzing an experiment, one can enjoy the advantages of relative states without having to worry about parallel worlds. I called it 'pocket addition' of the Everett theory.

In discussion with professor Dieter Zeh it became clear to me that if one assumes with Everett that there is no limit to the validity of \qm , i. e.,, that it is universally valid, then the many-worlds interpretation has no alternative in the approach.\\

{\bf (iv)} Authors of entertaining films \cite{Everettfilm} or television series caught on the idea of parallel worlds. In an episode of the television series Stargate the main character Sam Carter meets her replicas from parallel worlds as a person from Europe meets one from America. In a film I saw that meeting a replica led to explosive disappearance  of both.

These ideas are false, but understandable because a coherent sum of \Q components (as (1) or (2)) has no parallel in the classical world. Two components can be united into one, and the new state may be very different from both former components. (One should only think of the well-known two-slit interference pattern.)\\

{\bf (v)} The reader may wonder why stick to {\bf unitary dynamics} when there are other ways to complete pre\M (some of them presented in this volume). I admit that it is a hypothesis, a conviction, an act of faith. (Read perhaps my unpublished essay \cite{FHARXunit.dyn} and Schlosshower's scholarly study \cite{Schlossh}.)

The natural conceptual framework for Newtonian mechanics is the three-dimensional real space and the one-dimensional real time axis. This 'naturalness' is, of course, imprinted onto us biologically: our classical intuition 'feels' it.

Some of us believe that the complex Hilbert space with its unitary metrics, the state space,  is the natural conceptual framework of \qm . But it is not impressed on us by the still lacking \Q intuition. Unitary operators are isometries of the state space, and thus the unique way to formulate changes in the state space. On the other hand, unitary evolution operators are the integral equivalents of the famous (and elementary) Schr\"odinger equation.\\

{\bf (vi)} Bell has disproved \cite{Bell} von Neumann (physically, not mathematically): von Neumann's no-hidden-variables no-go theorem \cite{vNeum} (chapter IV, section 2 there). Similarly, {\bf Everett has disproved} von Neumann's famous claim that process 1 (complete \m ) cannot follow from process 2 (pre\m ) within unitary dynamics (cf Concluding remark (II) in section X in \cite{FHMeasAlg}).\\

{\bf (vii)} One should also take into account the recent ontic breakthrough theorem of Pusey, Barrett, and Rudolph \cite{PBR}, which has made a great splash in quantum foundations \cite{FHARX1}, \cite{Leifer} (perhaps see also my unpublished essay on historical background to the breakthrough \cite{FHARX2}). It can be viewed to give {\bf support} to the Everettian many-worlds relative-state interpretation of \qm . Here are the own words of the authors (near the end of their article).

"Finally, what are the consequences if we simply accept both the assumptions and the conclusion of the theorem? If the quantum state is a physical property of a system then quantum collapse must correspond to a — problematic and poorly defined — physical process. If there is no collapse, on the other
hand, then after a measurement takes place the joint quantum state of the system and measuring apparatus is entangled and contains a component corresponding to each possible macroscopic measurement outcome. This would be unproblematic if the
quantum state merely reflected a lack of information about which outcome occurred. However, if the quantum state is a physical property of the system and apparatus, it is hard to avoid the
conclusion that each macroscopically different component has a direct counterpart in reality."\\

{\bf (viii)} Quantum Mechanics needs to heal its almost a century long wound called the paradox of \M theory. It cannot perpetuate completing pre\M by some form of empirical information - like von Neumann's {\it ad hoc} projection postulate ("process 1") or Bohr's arbitrary invoking classical physics - because this would not be 'healing' the wound; it would be 'living with it' as it is done so far.

Quantum Mechanics seems to be at the crossroads.
It can take one of the following two 'roads'.
It can go beyond the unitary formalism, as do different interpretations (some of them presented in this volume), or it can remain within unitary dynamics and accept an Everettian many-worlds interpretation \cite{De Witt1}, \cite{newBOOK}, \cite{Wallace}.

The latter is "extravagant" according to one of the deepest thinkers in \Q foundations, the late John Stuart Bell. I would add that the many-worlds interpretation is disquieting, frightening, appalling; but it still can be the right 'road' to take.

A long time ago earth was believed to be the center of the universe (following Ptolemy). Can we imagine what amazement and confusion it was for the people of that time to imagine earth as one of the numerous moving celestial bodies (after Copernicus)?

Our descendants may find the idea of belonging to one of the many worlds as natural as we find the motion of the earth. They will then most likely have a well developed \Q intuition and perhaps they will wonder in astonishment why was the idea of parallel worlds  mind-boggling for us.\\

\appendix{\bf Appenix A. Elements of unitary premeasurement theory}

\vspace{4mm}
To begin with, the basic assumptions are shortly sketched and the notation is explained.

The subsystem that is measured is denoted by A. It is assumed that an observable \$O_A=\sum_ko_kE_A^k\$ with a finite or infinite purely discrete spectrum \$\{o_k:\forall k\}\$ is given in unique spectral form (by "uniqueness" is meant the fact that there is no repetition in the eigenvalues in the spectral form). Each eigenvalue \$o_k\$ can be arbitrarily degenerate.

The measuring instrument is denoted as subsystem B. The \M results are expressed in terms of the 'pointer positions', i. e., the eigen-projectors \$F_B^k\$ of the pointer observable \$P_B=\sum_kp_kF_B^k\$, which is accompanied by the completeness relation \$\sum_kF_B^k=I_B\$, where \$I_B\$ is the identity operator for subsystem B. The pre\M process starts at an initial moment \$t_\textbf{i}\$, when an initial pure state (state vector) \$\ket{\phi}_A\IN\$ and the pure initial (or 'ready-to-measure') state \$\ket{\phi}_B^\mathbf i\$ of the instrument are given. The interaction between object and apparatus is incorporated in the unitary evolution operator \$U_{AB}\equiv U_{AB}(t_\textbf{f}-t_\textbf{i})\$, which brings about the final (composite) state $$\ket{\Phi}_{AB}^\mathbf f\equiv U_{AB}\Big(\ket{\phi}_A^\mathbf i\ket{\phi}_B^\mathbf i\Big).\eqno{(A.1)}$$

Pre\M is defined  \cite{BLM} by the {\bf \CC } demanding that whenever it is statistically certain that the initial state \$\ket{\phi_A}^\mathbf i\$ has a sharp value \$o_{\bar k}\$ of the measured observable, then it must be statistically certain that the final composite state \$\ket{\Phi}_{AB}^\mathbf f\$ has the corresponding sharp value \$p_{\bar k}\$ of the pointer observable. (By "corresponding" is meant "having the same value of the index k", cf the two spectral forms above.) One can write this as
$$E_A^{\bar k}\ket{\phi}_A\IN =\ket{\phi}_A\IN\quad
\Rightarrow\quad
F_B^{\bar k}\ket{\Phi}_{AB}\f =\ket{\Phi}_{AB}\f $$
(cf relations (5) and (6) in \cite{FHMeasAlg}).
The symbol "$\Rightarrow$" denotes logical implication.

We are interested in pre\M theory that is confined to discrete ordinary observables. (For  more general observables see \cite{BLM}.) The \CC is {\bf equivalent} to the so-called {\bf \prc } for them, which requires the equality of the predicted probability of the eigenvalues of the measured observable in any initial state \$\ket{\phi}_A^\mathbf i\$ and that of the corresponding 'pointer positions' in the final composite state (1):
$$\forall \ket{\phi}_A\IN ,\enskip\forall k:\quad\bra{\phi}_A\IN E_A^k\ket{\phi}_A\IN\textbf{=}
\bra{\Phi}_{AB}\f F_B^k\ket{\Phi}_{AB}\f . \eqno{(A.2)}$$

We will need also a third equivalent {\bf dynamical definition} of pre\M , which reads: $$\forall \ket{\phi}_A\IN ,\enskip\forall k:\quad
F_B^kU_{AB}\Big(\ket{\phi}_A^\mathbf i\ket{\phi}_B^\mathbf i\Big)=U_{AB}E_A^k
\Big(\ket{\phi}_A^\mathbf i\ket{\phi}_B^\mathbf i\Big).\eqno{(A.3)}$$

Equivalence of the dynamical condition and the \CC is proved in \cite{FHSubsMeas}.\\

Since \$\sum_kF_B^k=I_B\$ and \$||F_B^k\ket{\Phi}_{AB}\f || =\Big(\bra{\Phi}_{AB}\f F_B^k\ket{\Phi}_{AB}\f\Big)^{1/2}\$, the \PRC (2) implies the decomposition

$$\ket{\Phi}_{AB}\f =\sum_k(\bra{\phi}_A\IN
E_A^k\ket{\phi}_A\IN )^{1/2}
\times\Big( F_B^k\ket{\Phi}_{AB}\f \Big/
||F_B^k\ket{\Phi}_{AB}\f ||\Big)\eqno{(A.4a)}$$ valid for every \$\ket{\phi}_A\IN\in \cH_A\$.

Expansion (A.4a) displays a {\bf connection between pre\M and complete \m }. Namely, the LHS of (A.4a) is the final state of pre\M of the given observable \$O_A=\sum_ko_kE_A^k\$,
whereas each term of the RHS of (A.4a) applies to one complete \m .

One should note that the complete-\M final states
$$\forall k,\enskip \bra{\phi}_A\IN E_A^k\ket{\phi}_A\IN  >0:\quad F_B^k\ket{\Phi}_{AB}\f \Big/||F_B^k\ket{\Phi}_{AB}\f ||\eqno{(A.4b)}$$ are based on the tacit assumption of {\bf minimal pre\M } \cite{FHminimal}, i. e., lack of any over\M is assumed. The latter would be the case if a finer observable, of which the measured observable \$O_A\$ is a non-trivial function, were measured implying the pre\M of \$O_A\$ (cf section V in \cite{FHMeasAlg}).\\

The {\bf dynamical definition (A.3) implies} that {\bf only} the \$k$-th initial component \$E_A^k
\ket{\phi}_A^\mathbf i\$ of the initial state \$\ket{\phi}_A\IN =\sum_kE_A^k \ket{\phi}_A\IN\$
(decomposition due to \$\sum_kE_A^k=I_A\$)
contributes to the \$k$-th final complete-\M component in the  unitary evolution of pre\M : $$\forall\ket{\phi}_A \IN\in \cH_A,\enskip\forall k:\quad F_B^k\ket{\Phi}_{AB}\f =
U_{AB}\Big(E_A^k\ket{\phi}_A\IN \otimes
\ket{\phi}_B\IN \Big).\eqno{(A.5)}$$ Relation (A.5) is actually (A.3) rewritten.

Claim (A.5) is relevant for complete \M of the observable \$O_A\$ in which the result \$o_k\$ is obtained because, as it was stated, this process ends in the state \$F_B^k\ket{\Phi}_{AB}\f \Big/
||F_B^k\ket{\Phi}_{AB}\f ||\$.\\

The simplest pre\M is {\bf ideal pre\m }. It can be defined in three equivalent ways:

{\bf (I)}  $$\forall  \ket{\phi}_A\IN\in\cH_A:\quad \ket{\Phi}_{AB}^\mathbf f =
\sum_k(E_A^k\ket{\phi}_A\IN )
\otimes\ket{\phi}_B^k,\eqno{(A.6a)}$$ where \$\{\ket{\phi}_B^k:\forall k\}\$ is, in general, an ortho-normal eigen-sub-basis of \$P_B\$ in \$\cH_B\$: \$\forall k:\enskip F_B^k\ket{\phi}_B^k=
\ket{\phi}_B^k\$.

{\bf (II)} The definition that ensues  may be called that of {\bf L\"uder's change-of -state}. It says that that the general final state \$\rho_A\f\$ of the object in ideal pre\M of an observable \$O_A\enskip\Big(=\sum_ko_kE_A^k\Big)\$
is given by: $$\forall\ket{\phi}_A\IN :\enskip \rho_A\f\equiv\tr_B\Big(\ket{\Phi}_{AB}\f \bra{\Phi}_{AB}\f\Big)
=\sum_kE_A^k\ket{\phi}_A\IN \bra{\phi}_A\IN E_A^k.\eqno{(A.6b)}$$

{\bf (III)} The third definition reads: Every initial state that has a sharp value of the measured observable does not change at all in ideal pre\M : $$\ket{\phi}_A\IN =E_A^{\bar k}\ket{\phi}_A\IN\quad\Rightarrow\quad
\rho_A\f=\ket{\phi}_A\IN \bra{\phi}_A\IN .\eqno{(A.6c)}$$An in-circle proof of the claimed equivalences is given in \cite{FHMeasAlg} (section VII there).\\

Finally, we shall make use of a general auxiliary claim, which will be referred to as the {\bf 'under-the-partial-trace commutativity'}. It reads:
$$Z_A\equiv\tr_B\Big(\mathbf{Y_B}X_{AB}\Big)=
\tr_B\Big(X_{AB}\mathbf{Y_B}\Big) ,\eqno{(A.7)}$$ where \$Y_B\$ and \$X_{AB}\$ are arbitrary subsystem and composite-system operators respectively. (Under the partial trace, by \$Y_B\$ is  denoted actually \$I_A\otimes Y_B\$.) It is straightforward to prove the general claim (A.7).\\

{\bf\noindent  Appendix B. Proof of equivalence of partial-scalar product and partial-trace}\\

\noindent
We now prove the following {\it equivalence} (relation (7)): $$\ket{\phi}_A=\bra{\phi}_B\ket{\Psi}_{AB}
\Big/[\bra{\Psi}_{AB}(\ket{\phi}_B \bra{\phi}_B)\ket{\Psi}_{AB}]^{1/2}
\quad\Leftrightarrow$$
$$\ket{\phi}_A \bra{\phi}_A=\tr_B[(\ket{\Psi}_{AB} \bra{\Psi}_{AB})(\ket{\phi}_B \bra{\phi}_B)]\Big/$$ $$\{\tr[(\ket{\Psi}_{AB} \bra{\Psi}_{AB})(\ket{\phi}_B \bra{\phi}_B)]\}.\eqno{(B.1)}$$
(Naturally, the implication "$\Leftarrow$" is 'up to an arbitrary phase factor'.)

Let us start assuming that \$\ket{\phi}_A\$ is given by the partial-scalar-product relation:
$$\ket{\phi}_A\bra{\phi}_A=
\Big(\bra{\phi}_B\ket{\Psi}_{AB}\Big)
\Big(\bra{\Psi}_{AB}\ket{\phi}_B
\Big)\Big/$$ $$\Big(\bra{\Psi}_{AB}(I_A\otimes\ket{\phi}_B
\bra{\phi}_B)\ket{\Psi}_{AB}\Big).\eqno{(B.2)}$$  First we evaluate the operator \$\cN_A\$ that is the nominator on the RHS of (B.2) utilizing two arbitrary complete orthonormal bases in the tensor-factor spaces \$\{\ket{i}_A:\forall i\},\enskip\{\ket{n}_B:\forall n\}\$.

$$\bra{i}_A\cN_A\ket{i'}_A=
\Big(\sum_n\bra{\phi}_B\ket{n}_B\bra{i}_A
\bra{n}_B\ket{\Psi}_{AB}\Big)\times
\Big(\sum_{n'}\bra{\Psi}_{AB}\ket{i'}_A
\ket{n'}_B\bra{n'}_B
\ket{\phi}_B\Big).\eqno{(B.3)}$$

On the other hand, if we evaluate the operator \$\cN_A'\$ that is the nominator on the RHS of the partial-trace relation (B.1) in the same two bases, we obtain
$$\bra{i}_A\cN_A'\ket{i'}_A=\sum_n\sum_{n'}
\bra{i}_A\bra{n}_B\ket{\Psi}_{AB}
\bra{\Psi}_{AB}\ket{i'}_A\ket{n'}_B
\bra{n'}_B\ket{\phi}_B
\bra{\phi}_B\ket{n}_B.\eqno{(B.4)}$$

Obviously, the terms on the RHS of (B.3) and (B.4) are equal (product of the same four numbers in different order). Hence, \$\cN_A=\cN_A'\$.

Next, we turn to the numbers that are the denominators \$\cD\$ and \$\cD'\$ of the partial-scalar product relation (B.2) and the partial-trace relations on the RHS of (B.1) respectively ,  and we ascertain of their equality.

 $$\cD =\sum_{i,n,i',n'}\bra{\Psi}_{AB}
 \ket{i}_A\ket{n}_B\bra{i}_A\bra{n}_B
 \Big(I_A\otimes \ket{\phi}_B\bra{\phi}_B\Big)$$ $$\ket{i'}_A\ket{n'}_B
 \bra{i'}_A\bra{n'}_B\ket{\Psi}_{AB}=$$ $$\sum_{i,n,i',n'}\bra{\Psi}_{AB}
 \ket{i}_A\ket{n}_B\times\delta_{i,i'}
\times$$  $$
 \bra{n}_B\ket{\phi}_B\times\bra{\phi}_B
 \ket{n'}_B\times
 \bra{i'}_A\bra{n'}_B\ket{\Psi}_{AB}.\eqno{(B.5)}$$

On the other hand we have
$$\cD'=\sum_{i,n,i',n'}\bra{i}_A \bra{n}_B
\ket{\Psi}_{AB}\bra{\Psi}_{AB}\ket{i}_A\ket{n'}_B
\bra{n'}_B\ket{\phi}_B\bra{\phi}_B\ket{n}_B.\eqno{(B.6)}$$

Again we can see that the corresponding terms in (B.5) and (B.6) coincide when we take into account the Kronecker symbol in the former and we exchange the mute indices \$n\$ and \$n'\$ in the latter. This concludes the proof of (B.1).\\

\pagebreak
{\bf\noindent  Appendix C. Complete ideal measurement in a general state}\\

As it is well known, ideal pre\M cannot be, in general, performed in direct pre\M (cf last passage in section VII. in \cite{FHMeasAlg}). This might seem to cast doubt on the significance of relation (8).

However, it was shown \cite{FHSubsMeas} that any general  pre\M of a subsystem event \$P_B\$ in any pure composite-system state \$\ket{\Psi}_{AB}\$ has the same effect on the opposite subsystem A as ideal pre\M , which is: $$\rho_A^{(Psi)}\equiv\tr_B\Big(P_B\ket{\Psi}_{AB}
\bra{\Psi}_{AB}P_B\Big)\Big/$$ $$\Big[\tr\Big(P_B\ket{\Psi}_{AB}
\bra{\Psi}_{AB}P_B\Big)\Big]\eqno{(C.1)}$$ (cf (33b) in \cite{FHMeasAlg}). One should note that the composite-system state (density operator) under the partial trace on RHS(C.1) is the state of those individual systems on which \$P_B\$ did occur (on some it did not).

Let \$\rho_{AB}\$ be an arbitrary (mixed or pure) composite-system state (density operator), and let $$\rho_{AB}=\sum_{k=1}^Kw_k\ket{\Psi}_{AB}^k
\bra{\Psi}_{AB}^k\eqno{(C.2)}$$ be a decomposition of the LHS into pure states. It is always possible. (A well-known way is the eigen-decomposition.)

I believe that Everett has followed Bohr in endowing the pure \Q state with twofold physical meaning: that of an ensemble of equally prepared systems and that of the individual systems making up the ensemble. This stipulation is made also in this study.

Since the pre\M process of \$P_B\$ in an ensemble described by \$\rho_{AB}\$ takes place on each individual system in the ensemble, and the ensemble is a mixture of pure states, clearly, it is true also for a general state that any pre\M of \$P_B\$ gives rise to the same state on subsystem A as ideal pre\M of \$P_B\$. (One may consider the claim for a pure state as the 'folded' version, and that for a general state as the 'unfolded' version because the generalization is unique.)

All we have to do is to find out {\it how ideal pre\M of \$P_B\$ changes \$\rho_{AB}\$}. Then the reduced density operator is the sought state of subsystem A. We have to find out how the statistical weights \$w_k\$ have to be modified when one takes into account the fact that the event \$P_B\$ does not occur on all individual systems in the ensemble.\\

Let us turn to the standard physical meaning of mixtures.

The physical meaning of (C.2) is that it determines a possible manner of preparation of an ensemble of \Q systems in the state \$\rho_{AB}\$: One has to take a sequence of integers $$N_k,\enskip k=1,2,\dots ,K\eqno{(C.3)}$$ such that, defining \$N\equiv\sum_{k=1}^KN_k\$, one has \$\lim_{(N_k)\to\infty}N_k/N=w_k,\enskip k=1,2,\dots ,K\$. Then one prepares \$N_k\$ systems in the state \$\ket{\Psi}_{AB}^k\$ for each value of \$k\$ to obtain a sub-ensemble that represents empirically and ensemblewise this pure state. The union of all these subensembles is then the ensemble representing empirically \$\rho_{AB}\$ (cf (C.2)). It consists of the \$N\$
\Q systems.

One should note that in the limit the proportions \$N_1:N_2:\dots :N_K\$ become equal to the proportions \$w_1:w_2:\dots :w_K\$ (and thus (C.3) determines the statistical weights \$w_k\$ in (C.2), the sum of which is normalized to 1).

When an event \$P_B\$ is measured on each of the \$N_k\$ physical systems in the state \$\ket{\Psi}_{AB}\$, on some of the states the result \$1\$ will be obtained (the event 'occurs'), let their number be \$N'_k\$. On \$N''_k\enskip\Big(=N_k-N'_k\Big)\$ the result \$0\$ is obtained ('non-occurrence').

Thus, for each \$k\$ value, each of the \$N'_k\$ individual-system states $$(\ket{\Psi}'_{AB})^k(\bra{\Psi}'_{AB})^k\equiv (P_B\ket{\Psi}_{AB}^k
\bra{\Psi}_{AB}^kP_B\Big)\Big/$$  $$ \Big[\tr\Big(\ket{\Psi}_{AB}^k
\bra{\Psi}_{AB}^kP_B\Big)\Big]\eqno{(C.4a)}$$ gives rise to a relative state $$(\rho'_A)^k\equiv\tr_B[(\ket{\Psi}_{AB}')^k
(\bra{\Psi}'_{AB})^k]
\eqno{(C.4b)}$$ for subsystem A (cf (8)).

Let us denote by \$p'_k\$ the relation \$N'_k/N_k\$. When \$N_k\$ tends to infinity, \$p'_k$ tends to the probability \$\bra{\Psi}_{AB}^kP_B
\ket{\Psi}_{AB}^k\$ (as we know fom the statistical interpretation of probability):
$$\forall k:\quad p'_k=\bra{\Psi}_{AB}^kP_B
\ket{\Psi}_{AB}^k.\eqno{(C.5)}$$

When the event \$P_B\$ has occurred, the relative frequencies are
$$N'_k\Big/\sum_{k'}N'_{k'}=N_kp'_k\Big/
\sum_{k'}N_{k'}p'_{k'}=$$  $$(N_k/N)p'_k\Big/
\sum_{k'}[(N_{k'}/N)p'_{k'}].\eqno{(C.6)}$$

When all \$N_k\$ tend to infinity, then, taking into account (C.5),  we obtain the new statistical weights \$w'_k\$ as the limit value of the last expression in (C.6):
$$w'_k=w_k\times\Big(\bra{\Psi}_{AB}^kP_B
\ket{\Psi}_{AB}^k\Big)\Big/$$ $$
\Big[\sum_{k'}w_{k'}\Big(
\bra{\Psi}_{AB}^{k'}P_B
\ket{\Psi}_{AB}^{k'}\Big)\Big].$$
When we replace here \$\Big(
\bra{\Psi}_{AB}^qP_B
\ket{\Psi}_{AB}^q\Big),\$ by the equivalent trace expression
\$\tr\Big[\Big(\ket{\Psi}_{AB}^q\bra{\Psi}_{AB}^q\Big)
P_B\Big]\$, \$q=k,k'\$, then we obtain

$$\forall k:\quad w'_k=w_k\times
\tr\Big[\Big(\ket{\Psi}_{AB}^k\bra{\Psi}_{AB}^k\Big)
P_B\Big]\times$$
$$\Big\{\sum_{k'}w_{k'}\times
\tr\Big[\Big(\ket{\Psi}_{AB}^{k'}\bra{\Psi}_{AB}^{k'}
\Big)P_B\Big]\Big\}^{-1}.$$

In view of (C.2), this gives
$$\forall k:\quad w'_k=w_k\times
\tr\Big[\Big(\ket{\Psi}_{AB}^k\bra{\Psi}_{AB}^k\Big)
P_B\Big]\times$$ $$\Big[\sum_{k'}\tr\Big(\rho_{AB}
P_B\Big)\Big]^{-1}.\eqno{(C.7)}$$

Denoting by \$\rho_A\$ the state of subsystem A that comes about when \$P_B\$ is measured ideally in the general state \$\rho_{AB}\$ (cf (C.2) and (C4b)), we have $$\rho_A=\sum_kw'_k(\rho'_A)^k.$$ Substituting here \$w'_k\$ from (C.7) and \$(\rho'_A)^k\$ from (C.4b) and (C.4a), we finally obtain $$\rho_A=\tr_B\Big(\rho_{AB}P_B\Big)\Big/
\Big[\tr
\Big(\rho_{AB}P_B\Big)\Big].\eqno{(C.8a)}$$
This is equivalent to (cf (AUX-c) and (AUX-d) in section 4) $$\rho_A=\tr_B\Big(P_B\rho_{AB}P_B\Big)\Big/
\Big[\tr
\Big(P_B\rho_{AB}P_B\Big)\Big]\eqno{(C.8b)}$$ as it was claimed .\\

\pagebreak

\end{document}